\documentclass[fleqn,usenatbib]{mnras}

\usepackage{newtxtext,newtxmath}
\usepackage{comment}

\usepackage[T1]{fontenc}

\DeclareRobustCommand{\VAN}[3]{#2}
\let\VANthebibliography\thebibliography
\def\thebibliography{\DeclareRobustCommand{\VAN}[3]{##3}\VANthebibliography}

\usepackage{graphicx}	
\usepackage{amsmath}	
\usepackage{xspace}

\def\Euclid{\textit{Euclid}\xspace}

\title[$m-$bias reconstruction]{Reconstructing spatially-varying multiplicative bias for Stage\,IV weak lensing galaxy surveys with a quadratic estimator}

\author[K. Tanidis et al.]{
Konstantinos Tanidis,$^{1,2}$\thanks{E-mail: konstantinos.tanidis@unito.it}
David Alonso,$^{1}$
Lance Miller$^{1}$
and Joachim Harnois-D\'eraps$^{3}$
\\
$^{1}$Department of Physics, University of Oxford, Denys Wilkinson Building, Keble Road, Oxford OX1 3RH, UK\\
$^{2}$Center for Astrophysics and Cosmology, University of Nova Gorica,1280
Nova Gorica, Slovenia\\
$^{3}$School of Mathematics, Statistics and Physics, Newcastle University, Newcastle-upon-Tyne, NE1 7RU, UK
}

\date{Accepted XXX. Received YYY; in original form ZZZ}

\pubyear{\the\year{}}

\begin{document}
\label{firstpage}
\pagerange{\pageref{firstpage}--\pageref{lastpage}}
\maketitle

\begin{abstract}
  We present a quadratic estimator that detects and reconstructs spatially-varying multiplicative ($m-$) bias in weak lensing shear measurements, by exploiting the $EB$ mode coupling that it generates. The method combines $E$ and $B$ modes with inverse-variance weights, to yield an unbiased reconstruction of $m(\boldsymbol{\theta})$ to first order. We study the ability of future Stage\,IV surveys to obtain an unbiased reconstruction of the $m$-bias in differing scenarios, considering differing bias morphologies, and characteristic scales, as well as differing metrics to quantify the signal-to-noise ratio of the reconstructed  map. Considering an $m$ pattern repeating on $\sim 1^\circ\times1^\circ$ sky patches, as might be the case for an $m$ field caused by focal-plane systematics. With a \Euclid-like redshift distribution, we find that $\sim5\%$ rms variations in $m$-bias may be detected at the 20$\sigma$ level, after stacking between $\sim400$ and $\sim1000$ patches (rising to between $\sim2800$ and $\sim7600$ for $1\%$ rms variations, data volumes that are becoming available with upcoming surveys), depending on the morphology of the $m$ pattern. We show that these results are robust against the cosmological model assumed in the reconstruction, as well as the presence of intrinsic alignments or baryonic effects, and that the method shows no spurious response to additive ($c-$) bias. These results demonstrate that percent-level, spatially-varying $m-$bias can be detected at high significance, enabling diagnosis and mitigation in the Stage\,IV weak lensing era.
\end{abstract}




\section{Introduction}
\label{sec:intro}
Stage\,IV weak lensing surveys have the potential to make significant improvements in our understanding of the cosmological model \citep{albrecht_2006}, a prospect which is now being realised with large-area, high-accuracy surveys such as \Euclid \citep{mellier_2025} and the Vera Rubin Observatory’s Legacy Survey of Space and Time \citep[LSST:][]{lsstref}.  However, it is widely recognised that the achievement of accurate cosmological inference from weak lensing surveys requires understanding and mitigation of a wide range of systematic effects \citep{mandelbaum_2018}.

In the weak lensing regime, where the shear due to gravitational lensing is small (\la 0.01), it is common to express the effects of systematic biases to leading order as linear multiplicative ($m-$) and additive ($c-$) biases \citep{huterer_2006, heymans_2006},
\begin{equation}
\label{eq:reduced_shear}
\gamma^{\rm biased} \simeq (1+m) \gamma^{\rm cosm} + c \, ,
\end{equation}
for biased and true cosmic shear, $\gamma^{\rm biased}$ and $\gamma^{\rm cosm}$ respectively\footnote{We should note that here we neglect the distinction between reduced shear $g$ and shear $\gamma$, as it does not make a difference for the formalism we develop in this work. Therefore, from now on, we treat $\gamma^{\rm biased}$ as the observed shear field.}. There may be multiple effects contributing to these biases, including both astrophysical and measurement systematics.  Among the set of measurement systematics, the errors in the modelling and correction for the imaging point spread function (PSF) may cause both $m-$ and $c-$biases \citep{paulin_henriksson_2008, massey_2013}, and these biases may be further decomposed into correlations with parameters of the measurement, such as “PSF leakage” \citep{jarvis_2016}.

Among the additive biases, methods have been developed to detect these in data. One set of methods is based around cross-correlation with expected sources of error, such as the “Rho” statistics of \citet{rowe_2010} and \citet{jarvis_2016}, and “Tau” statistics of \citet{gatti_2021}, with cross-correlation between measured galaxy shapes and both PSF ellipticity and size and PSF modelling error.  Alternatively, additive bias that has some repeating pattern, for example due to time-invariant errors on the scale of the field-of-view (FOV) due to PSF modelling error, may be revealed in the data by stacking of galaxy ellipticity measures from a wide sky area in FOV coordinates \citep[e.g.][]{van_Uitert_2016, hildebrandt_2020}. Finally, template deprojection techniques may be employed to mitigate the impact of additive bias from sources with a known spatial pattern \citep{Alonso_2019, cornish2025systematicsmitigationcataloguebasedangular}.

PSF modelling errors may also generate multiplicative bias: specifically, it is expected that errors in the knowledge of PSF size lead to joint $m-$ and $c-$bias \citep{paulin_henriksson_2008, massey_2013}. The statistical measurement of galaxy shear generally also introduces $m-$bias \citep{refregier_2012, melchior_2012}, which must be corrected either by realistic simulations  \citep[e.g.][]{li_2023} or by self-calibration \citep[e.g. "metacalibration",][]{huff_2017, Sheldon_2017}. Further sources of measurement $m-$bias include the effects of galaxy neighbours, blends and redshift biases \citep{samuroff_2018, maccrann_2022}. However, the direct detection of $m-$bias in weak lensing data is problematic: the shear-ratio test of \citet{jain_2003} and \citet{taylor_2007} may be used to detect redshift-dependent $m-$bias \citep{schneider_2016},  but only with limited sensitivity.  For this reason, obtaining accurate knowledge of the total $m-$bias in Stage\,IV weak lensing surveys is considered to be one of the most challenging aspects of the analysis \citep{mandelbaum_2018}.

When calibrating $m-$bias in Stage\,III surveys, from simulations, it has generally been assumed that the bias may be dependent on galaxy properties, such as galaxy size, and hence may also be redshift-dependent \citep{li_2023}. However, it has been assumed that the bias is otherwise invariant with sky position. It had been thought that the effects of spatially varying $m-$bias would be negligible, with accurate knowledge only required of the mean $m-$bias \citep{kitching_2019}. More recently, however, it has been recognised that spatially-dependent $m-$bias may have a significant convolutional effect on measured shear power spectra: for Stage\,IV surveys, percent-level rms variations in $m-$bias on angular scales of degrees or larger may introduce a significant error in Stage\,IV cosmology analyses \citep{cragg_2023}.

A spatially-varying $m-$bias effectively leads to a spatial modulation of the true shear field, inducing additional statistical couplings between different independent Fourier modes and shear components. This is reminiscent of the impact of gravitational lensing on the Cosmic Microwave Background (CMB) temperature and polarisation anisotropies, in the form of a spatial modulation proportional to the gradient of these anisotropies. The statistical coupling induced by this modulation can then be used to reconstruct the field causing it (the lensing potential or the reionization in the case of CMB lensing, or the $m-$bias in the case of cosmic shear), by considering correlations between unequal Fourier modes, through the so-called ``quadratic estimator'' approach \citep{Lewis_2006, Dvorkin_2009}. Other applications of the quadratic estimator include the removal of CMB systematics \citep{Williams_2021}. Here we present a quadratic estimator for the reconstruction and detection of a spatially-dependent $m-$bias in cosmic shear data.

This paper is structured as follows. Section 2 introduces the method. We discuss how the method may be applied either to cross-correlation with templates of potential systematic effects, or in a direct-detection method where we stack on the scale of the FOV – in either case somewhat analogously to what may already be achieved for additive bias detection and measurement. Section 3 describes how the method is demonstrated in this paper, using some simplified models, with results and conclusions in sections 4 and 5.

\section{Method}
\subsection{Introduction to the method}
\label{subsec: intro_to_method}
The $m-$bias detection method acts on the harmonic space 2-point functions, $E$ and $B$, which we start by defining here.
For the demonstration in this paper,
we assume the flat-sky approximation, so that angular coordinates $\boldsymbol{\theta}$ can be treated as two-dimensional vectors on a plane\footnote{For analysis on large angular scales, a curved-sky methodology could be adopted \citep[e.g. ][]{Dvorkin_2009}.}. The spin-2 cosmic shear field is described by two components $\gamma_1(\boldsymbol{\theta})$ and $\gamma_2(\boldsymbol{\theta})$, which can be combined into a complex field $\gamma(\boldsymbol{\theta}) = \gamma_1(\boldsymbol{\theta}) + i\gamma_2(\boldsymbol{\theta})$. This field can be decomposed into $E$-mode and $B$-mode components via the Fourier transform of $\gamma$:
\begin{equation}
\gamma(\boldsymbol{\theta}) = \int \frac{d^2\boldsymbol{\ell}}{(2\pi)^2}{\gamma}(\boldsymbol{\ell}) e^{i\boldsymbol{\ell}\cdot\boldsymbol{\theta}} \, ,
\end{equation}
where ${\gamma}(\boldsymbol{\ell})$ is the shear in Fourier-space. In terms of the Fourier-space $\gamma_1$ and $\gamma_2$, one can obtain the $E$ and $B$ modes by a rotation in Fourier space (e.g. \citealt{Hu_2000,Schneider_2006}):
\begin{eqnarray}
E(\boldsymbol{\ell}) = {\gamma}_1(\boldsymbol{\ell})\cos(2\varphi_{\boldsymbol{\ell}}) + {\gamma}_2(\boldsymbol{\ell})\sin(2\varphi_{\boldsymbol{\ell}}), \label{eq:E-mode-def}\\
B(\boldsymbol{\ell}) = -{\gamma}_1(\boldsymbol{\ell})\sin(2\varphi_{\boldsymbol{\ell}}) + {\gamma}_2(\boldsymbol{\ell})\cos(2\varphi_{\boldsymbol{\ell}}) \, , 
\label{eq:B-mode-def}
\end{eqnarray}
with $\varphi_{\boldsymbol{\ell}}$ the polar angle of the wavevector $\boldsymbol{\ell}$. By construction, $E(\boldsymbol{\ell})$ and $B(\boldsymbol{\ell})$ are real-valued fields describing the parity-even and parity-odd components of the shear field, respectively. At linear order, and in the absence of systematics, the cosmological shear signal $\gamma^{\rm cosm}$, due to gravitational lensing by large-scale structure, is expected to produce entirely $E$-mode power. We therefore consider the cosmological $B$-mode to be zero, $B^{\rm cosm}(\boldsymbol{\ell}) = 0$, and that all cosmological shear information resides in $E^{\rm cosm}(\boldsymbol{\ell})$\footnote{Although this is a good approximation in linear order, there are also non-linear lensing effects producing small contributions to $B$-modes \citep{Schneider_2006b, Saga_2015}. It is also important to note that intrinsic alignments can also produce $B$-modes \citep{Blazek_2019}.}. Assuming statistical isotropy, the $E$-mode angular power spectrum $C_{\ell}^{EE}$ is defined by
\begin{equation}
  \langle E^{\rm cosm}(\boldsymbol{\ell}) E^{{\rm cosm}*}(\boldsymbol{\ell}') \rangle = (2\pi)^2 \delta^{\text{D}}(\boldsymbol{\ell}-\boldsymbol{\ell}') C_{\ell}^{EE},
\end{equation}
where $\delta^{\text{D}}$ is the two-dimensional Dirac delta function. Thus, any measured $B$-modes in the observed shear field $\gamma^{\rm biased}$ could therefore be an indicator of systematic effects or noise. In the current work, we focus on building an estimator which detects the mode-coupling between $E$ and $B$, as well as between Fourier modes with different wave vectors, as induced by a spatially-varying multiplicative bias field contaminating the cosmic shear signal.

\subsection{Spatially-varying shear bias and mode-coupling}
\label{subsec:spatially-varying systematics}
We model a spatially-varying multiplicative shear bias $m(\boldsymbol{\theta})$ as a dimensionless scalar field that modulates the true cosmological shear. Eq.\,\ref{eq:reduced_shear} becomes:
\begin{equation}\label{eq:mul_bias_real}
\gamma^{\rm biased}(\boldsymbol{\theta}) \simeq \big[1 + m(\boldsymbol{\theta})\big]\gamma^{\rm cosm}(\boldsymbol{\theta})+c(\boldsymbol{\theta}) \, ,
\end{equation}
where $c(\boldsymbol{\theta})$ is also a spatially-varying $c-$bias contamination. In practice, $m\ll 1$ and $c\ll1$ for weak systematics. In Fourier space, the product in Eq.\,\ref{eq:mul_bias_real} becomes a convolution. Let ${\gamma}^{\rm biased}(\boldsymbol{\ell})$ and ${\gamma}^{\rm cosm}(\boldsymbol{\ell})$ be the Fourier transforms of $\gamma^{\rm biased}(\boldsymbol{\theta})$ and $\gamma^{\rm cosm}(\boldsymbol{\theta})$, respectively, and ${m}(\boldsymbol{L})$ and ${c}(\boldsymbol{\ell})$ the Fourier transforms of the $m-$bias and the $c-$bias fields. Then, we have:
\begin{equation}\label{eq:conv_fourier}
{\gamma}^{\rm biased}(\boldsymbol{\ell}) \simeq {\gamma}^{\rm cosm}(\boldsymbol{\ell}) + \int \frac{d^2\boldsymbol{L}}{(2\pi)^2}{m}(\boldsymbol{L})\,{\gamma}^{\rm cosm}(\boldsymbol{\ell}-\boldsymbol{L})+{c}(\boldsymbol{\ell}) \, .
\end{equation}
The first term is the cosmic shear at wavevector $\boldsymbol{\ell}$, the second term is the contribution from the coupling between the $m-$bias and the cosmological shear field and the third term is the $c-$bias. The coupling term encodes how $m(\boldsymbol{\theta})$ redistributes power in the shear field: it mixes Fourier modes separated by wavevector $\boldsymbol{L}$. In particular, the $m-$bias field produces off-diagonal correlations in Fourier space. Decomposing Eq.\,\ref{eq:conv_fourier} into $E$ and $B$ components using Eqs\,\ref{eq:E-mode-def}--\ref{eq:B-mode-def}, one finds that a non-zero $m-$bias field mixes $E$ and $B$ modes. Intuitively, a spatially-varying multiplicative factor introduces spurious ``curl'' into the shear pattern, sourcing $B$-modes from an originally pure $E$-mode field.

More concretely, inserting Eq.~(\ref{eq:conv_fourier}) into the definitions of $E^{\rm biased}$ and $B^{\rm biased}$, and assuming here $B^{\rm cosm}\ne0$ for generality, we can derive the first-order effect of $m-$bias on the observed $E$ and $B$ fields. The observed $E$-mode receives a direct contribution from the cosmological $E$ plus a convolution with $m-$bias and also a $c-$bias term:
\begin{multline}\label{eq:E_obs}
E^{\rm biased}(\boldsymbol{\ell}) = E^{\rm cosm}(\boldsymbol{\ell})\\
+ \int \frac{d^2\boldsymbol{L}}{(2\pi)^2}~m(\boldsymbol{L})\{E^{\rm cosm}(\boldsymbol{\ell}-\boldsymbol{L})\cos[2(\varphi_{\boldsymbol{\ell}-\boldsymbol{L}} - \varphi_{\boldsymbol{\ell}})]\\
-B^{\rm cosm}(\boldsymbol{\ell}-\boldsymbol{L})\sin[2(\varphi_{\boldsymbol{\ell}-\boldsymbol{L}} - \varphi_{\boldsymbol{\ell}})]\}+E_{\rm c}(\boldsymbol{\ell}) \, ,
\end{multline}
and the observed $B$-mode is similarly given by:

\begin{multline}\label{eq:B_obs}
B^{\rm biased}(\boldsymbol{\ell}) = B^{\rm cosm}(\boldsymbol{\ell})\\
+\int \frac{d^2\boldsymbol{L}}{(2\pi)^2}~m(\boldsymbol{L})\{E^{\rm cosm}(\boldsymbol{\ell}-\boldsymbol{L})\sin[2(\varphi_{\boldsymbol{\ell}-\boldsymbol{L}} - \varphi_{\boldsymbol{\ell}})]\\
+B^{\rm cosm}(\boldsymbol{\ell}-\boldsymbol{L})\cos[2(\varphi_{\boldsymbol{\ell}-\boldsymbol{L}} - \varphi_{\boldsymbol{\ell}})]\}+B_{\rm c}(\boldsymbol{\ell}) \, .
\end{multline}

Here, the $\cos(2\Delta\varphi)$ and $\sin(2\Delta\varphi)$ factors arise from the spin-2 nature of the shear field. An important consequence of Eqs\,\ref{eq:E_obs}--\ref{eq:B_obs} is that $E$ and $B$ modes become statistically correlated in the presence of $m-$bias, even if the $c-$bias is zero. In the absence of $m-$bias and $c-$bias, one would have $\langle E^{\rm biased}(\boldsymbol{\ell}) B^{\rm biased}(\boldsymbol{\ell}')\rangle=0$ (since $B^{\rm biased}$ would be pure noise or very small non-linear weak lensing cosmological signals, uncorrelated with $E^{\rm biased}$). With $m \neq 0$, the coupling induces a characteristic off-diagonal correlation between $E$ and $B$. To first order in $m$, using Eq.\,\ref{eq:conv_fourier} or directly from Eqs\,(\ref{eq:E_obs})–(\ref{eq:B_obs}), one finds for $\boldsymbol{\ell} - \boldsymbol{\ell}' = \boldsymbol{L}$ that
\begin{multline}\label{eq:EB_correlation}
\langle E^{\rm biased}(\boldsymbol{\ell}) B^{\rm biased}(\boldsymbol{\ell}-\boldsymbol{L}) \rangle = \\
\Big[ \Big(C_{\ell}^{EE} - C_{|\boldsymbol{\ell}-\boldsymbol{L}|}^{BB}\Big) \sin 2(\varphi_{\boldsymbol{\ell}} - \varphi_{\boldsymbol{\ell}-\boldsymbol{L}}) \Big]m(\boldsymbol{L})+E_{\rm c}(\boldsymbol{\ell})B_{\rm c}(\boldsymbol{\ell}-\boldsymbol{L}) \, .
\end{multline}

To linear order, we treat $m$ and $c$ as fixed external fields and average over the cosmic shear as a statistically isotropic field, while we neglect additional cross and higher-order terms. Equation~(\ref{eq:EB_correlation}) is the key statistical signature of a spatially-varying $m-$bias: The first term shows that an $E$-mode at wavevector $\boldsymbol{\ell}$ couples to a $B$-mode at wavevector $\boldsymbol{\ell}-\boldsymbol{L}$ such that their correlation is proportional to the bias field mode $m(\boldsymbol{L})$. The second term acts as a mean field $c-$bias contamination to the $m-$bias response, if this is present.

The methodology we develop here is directly analogous to the case of CMB lensing, where lensing by a potential $\phi$ induces correlations between $E$ and $B$ polarization modes of the CMB. In fact, Eq.\,\ref{eq:EB_correlation} is the weak lensing shear counterpart of the CMB lensing relation given in \cite{Lewis_2006} for $EB$ mode coupling by the lensing potential (see their Sec. 7.2).

\subsection{Quadratic estimator for the $m-$bias field}
\label{subsec:quadratic_estimator}
 Now, we construct a quadratic estimator for $m$ in Fourier space, denoted $\hat{m}(\boldsymbol{L})$, formed from the quadratic combination of observed $E$ and $B$ fields. Following the formalism of \cite{Lewis_2006}, we can write the Fourier-space estimator for the $m-$bias field as:
\begin{equation}\label{eq:estimator_def}
\hat{m}(\boldsymbol{L}) = \mathcal{N}(\boldsymbol{L}) \int \frac{d^2\boldsymbol{\ell}}{(2\pi)^2}
E^{\rm biased}(\boldsymbol{\ell})B^{\rm biased}(\boldsymbol{\ell}-\boldsymbol{L}) W(\boldsymbol{\ell},\boldsymbol{\ell}-\boldsymbol{L}) \, ,
\end{equation}
where $W(\boldsymbol{\ell},\boldsymbol{\ell}-\boldsymbol{L})$ is a weighting function to be specified, and $\mathcal{N}(\boldsymbol{L})$ is an overall normalization factor.  Eq.\ref{eq:EB_correlation} shows, if $c-$bias is zero, that the quadratic product $X_{\boldsymbol{\ell},{\bf L}}=E^{\rm biased}(\boldsymbol{\ell}) B^{\rm biased}(\boldsymbol{\ell}-\boldsymbol{L})$ has mean
\begin{equation}
  \langle X_{\boldsymbol{\ell},{\bf L}}\rangle=f(\boldsymbol{\ell}, \boldsymbol{\ell}-\mathbf{L})\,m(\mathbf{L}),
\end{equation}
where
\begin{equation}
  f(\boldsymbol{\ell} ,\boldsymbol{\ell}-\mathbf{L})=(C_{\ell}^{EE} - C_{|\boldsymbol{\ell}-\boldsymbol{L}|}^{BB}) \sin 2(\varphi_{\boldsymbol{\ell}} - \varphi_{\boldsymbol{\ell}-\boldsymbol{L}}).
\end{equation}
Therefore, we estimate $m(\mathbf L)$ by filtering $X_{\boldsymbol{\ell},{\bf L}}$ against the response $f$ with inverse-variance weights for $W$, and use $N$ to recover an unbiased, minimum-variance estimator. We note again here that, if there is a mean field $EB$ $c-$bias contamination in Eq.\ref{eq:EB_correlation}, this also enters the $m-$bias reconstruction in Eq.\ref{eq:estimator_def}. The integration runs over all Fourier modes $\boldsymbol{\ell}$, pairing an $E$-mode at $\boldsymbol{\ell}$ with a $B$-mode at $\boldsymbol{\ell}-\boldsymbol{L}$ so that the pair corresponds to the $m-$bias mode $\boldsymbol{L}$. Because the estimator is quadratic in the observed shear field (proportional to $E \times B$), it constitutes a quadratic estimator for $m$.

The role of $W(\boldsymbol{\ell}, \boldsymbol{\ell}-\boldsymbol{L})$ is to optimally weight each $(E,B)$ mode pair in the sum. Intuitively, modes with a higher signal-to-noise, either due to larger cosmological signal or lower noise, should receive higher weight. We derive the form of $W$ by demanding that $\hat{m}$ is the minimum-variance unbiased estimator of $m$. First, the unbiasedness condition means that the expectation value of $\hat{m}$ should equal the true $m$. However, until now, we have assumed that there is a bias term from the $c$ mean field. Thus, using Eq.\,\ref{eq:EB_correlation}, we require:
\begin{eqnarray}\label{eq:biased_condition}
\langle \hat{m}(\boldsymbol{L}) \rangle = m(\boldsymbol{L})+\mathcal{N}(\boldsymbol{L}) \int \frac{d^2\boldsymbol{\ell}}{(2\pi)^2}
E_{\rm c}(\boldsymbol{\ell})B_{\rm c}(\boldsymbol{\ell}-\boldsymbol{L}) W(\boldsymbol{\ell},\boldsymbol{\ell}-\boldsymbol{L}) \, .
\end{eqnarray}

At this point, we should note that the $c-$bias, as we have already discussed in Sec.\ref{sec:intro}, can be detected and, in principle, corrected using null tests from the data itself \citep[e.g.][]{van_Uitert_2016}. Assuming that the data we treat here from now on are $c-$bias corrected unless otherwise stated, Eq.\ref{eq:biased_condition} becomes $\langle \hat{m}(\boldsymbol{L}) \rangle = m(\boldsymbol{L})$. Then,  we can substitute the definition (\ref{eq:estimator_def}) and use $\langle E^{\rm biased}(\boldsymbol{\ell}) B^{\rm biased}(\boldsymbol{\ell}-\boldsymbol{L})\rangle$ from Eq.\,\ref{eq:EB_correlation}, so that the condition becomes
\begin{equation}\label{eq:unbiased_condition}
\mathcal{N}(\boldsymbol{L}) \int \frac{d^2\boldsymbol{\ell}}{(2\pi)^2}
\Big( C_{\ell}^{EE} - C_{|\boldsymbol{\ell}-\boldsymbol{L}|}^{BB} \Big)\sin\big[2(\varphi_{\boldsymbol{\ell}} - \varphi_{\boldsymbol{\ell}-\boldsymbol{L}})\big]
W(\boldsymbol{\ell},\boldsymbol{\ell}-\boldsymbol{L}) = 1 \, ,
\end{equation}
to linear order in $m$. Equation~(\ref{eq:unbiased_condition}) fixes the normalization factor $\mathcal{N}(\boldsymbol{L})$ in terms of the chosen weight function $W$. In practice, we  choose $W$ (up to an overall normalization) to minimize the variance of $\hat{m}(\boldsymbol{L})$. The variance $\mathrm{Var}[\hat{m}(\boldsymbol{L})]$ can be derived by considering $\langle \hat{m}(\boldsymbol{L}) \hat{m}^*(\boldsymbol{L}')\rangle$ and assuming Gaussian statistics for the cosmic shear field $E^{\rm cosm}(\boldsymbol{\ell})$. 
In this limit, we can use \citet{PhysRev.80.268}’s theorem to express the resulting four-point correlators of $E^{\rm cosm}$ into sums of products of 2-point functions. The dominant contributions to the estimator variance come from terms where the $E$ and $B$ fields in Eq.\,\ref{eq:estimator_def} contract with themselves, leading to noise bias. One finds that the variance is minimized if $W$ is chosen proportional to the signal coupling kernel from Eq.\,\ref{eq:EB_correlation}, divided by the total power in each of the $E$ and $B$ modes (signal + noise) \citep[c.f.][]{Lewis_2006}. In other words, the optimal weight is
\begin{equation}\label{eq:weight_function}
W(\boldsymbol{\ell}, \boldsymbol{\ell}-\boldsymbol{L}) = \frac{\displaystyle \sin\big[2(\varphi_{\boldsymbol{\ell}} - \varphi_{\boldsymbol{\ell}-\boldsymbol{L}})\big]}{\displaystyle C_{\ell}^{EE,\mathrm{tot}}C_{|\boldsymbol{\ell}-\boldsymbol{L}|}^{BB,\mathrm{tot}}}\Big( C_{\ell}^{EE} - C_{|\boldsymbol{\ell}-\boldsymbol{L}|}^{BB} \Big) \, ,
\end{equation}
where $C_{\ell}^{EE,\mathrm{tot}} = C_{\ell}^{EE} + N_{\ell}^{EE}$ and $C_{\ell}^{BB,\mathrm{tot}} = C_{\ell}^{BB} + N_{\ell}^{BB}$ denote the total $E$- and $B$-mode angular power spectra including noise\footnote{We note here that if $c-$bias is present, it contributes to the variance as $C_{\ell}^{XY,\mathrm{tot}} = C_{\ell}^{XY} + N_{\ell}^{XY} + C_{\ell}^{XY, \rm c}$.}. Here $N_{\ell}^{EE}$ and $N_{\ell}^{BB}$ are the noise power spectra of the $E$ and $B$ fields from intrinsic shape noise in the shear catalogs (note that shape noise typically contributes equally to $E$ and $B$ modes, on average). The factor $(C_{\ell}^{EE} - C_{|\boldsymbol{\ell}-\boldsymbol{L}|}^{BB})$ in the numerator of Eq.\,\ref{eq:weight_function} arises from the fact that either leg of the estimator pair could carry the signal coupling. The $\sin[2(\varphi_{\boldsymbol{\ell}} - \varphi_{\boldsymbol{\ell}-\boldsymbol{L}})]$ factor encodes the geometric dependence of the mode coupling (as in Eq.\,\ref{eq:EB_correlation}), ensuring that only the appropriate cross-orientation of $E$ and $B$ modes contributes to the signal. With this choice, modes are weighted by the inverse of their variance (the product of their total power spectra), which is the standard optimal weighting for a quadratic estimator.

Given the weight function (\ref{eq:weight_function}), we can now specify the normalization $\mathcal{N}(\boldsymbol{L})$ using Eq.\,\ref{eq:unbiased_condition}:
\begin{equation}\label{eq:normalization}
\mathcal{N}(\boldsymbol{L})^{-1} = \int \frac{d^2\boldsymbol{\ell}}{(2\pi)^2} \frac{\big(C_{\ell}^{EE} - C_{|\boldsymbol{\ell}-\boldsymbol{L}|}^{BB}\big)^2}{C_{\ell}^{EE,\mathrm{tot}}C_{|\boldsymbol{\ell}-\boldsymbol{L}|}^{BB,\mathrm{tot}}}\sin^2\big[2(\varphi_{\boldsymbol{\ell}} - \varphi_{\boldsymbol{\ell}-\boldsymbol{L}})\big] \, .
\end{equation}
At this point, we should mention that in the case of cosmic shear, and contrary to the CMB lensing case, the $m-$bias affects the measured shear which already includes the shape noise. Thus, it is worth checking the inclusion of the shape noise also in the numerators of Eqs\,\ref{eq:weight_function} and \ref{eq:normalization}, despite using theory predictions for the spectra. We confirmed that the performance of the quadratic estimator is not impacted, and decided to retain the original formalism, with the shape noise only in the denominator.

In practice, one can compute $\mathcal{N}(\boldsymbol{L})$ given a fiducial $E$-mode angular power spectrum, assuming a cosmological model and known noise levels. 

\begin{figure*}
    \centering
    \includegraphics[width=1.0\linewidth]{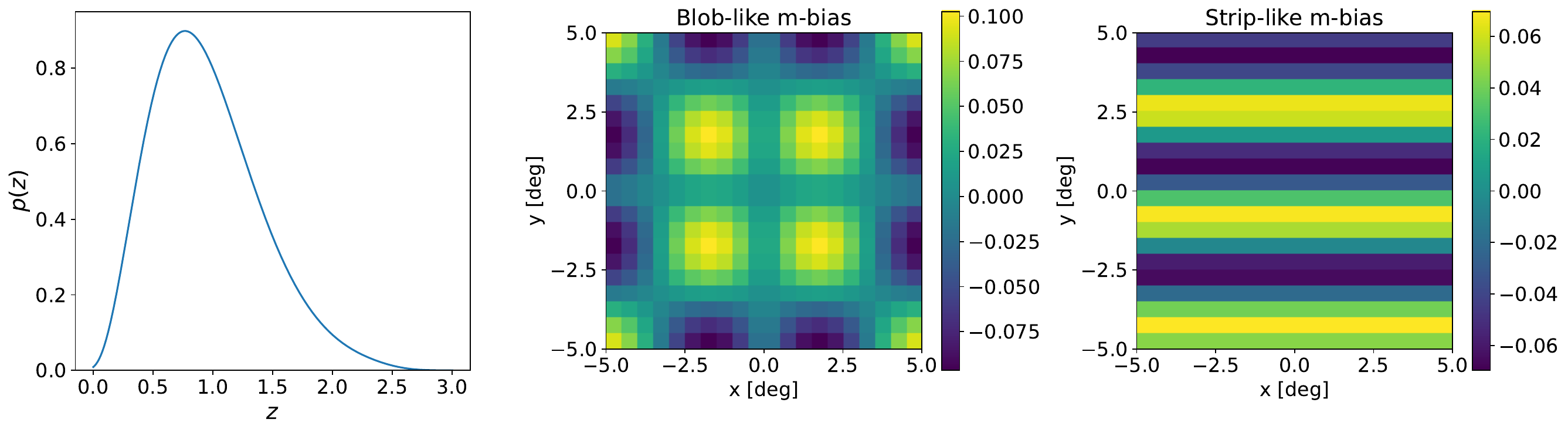}
    \caption{\textit{Left:} Galaxy redshift distribution $p(z)$ for a \Euclid-like photometric sample (see Eq.\ref{eq:n_of_z}). \textit{Middle:} A blob-like $m-$bias given by Eq.\ref{eq:blob_mbias}. \textit{Right}: A strip-like $m-$bias model given by Eq.\ref{eq:strip_mbias}. Both $m-$bias patterns have an rms of 5\% and are $20\times20$ pixel grids.}
    \label{fig:n_z_and_m_bias_input}
\end{figure*}

In principle, a more general estimator may be derived, using also the $EE$ and $BB$ correlations of the observed shear field, similar to the global minimum variance quadratic estimator used in CMB lensing reconstruction \citep{Maniyar_2021}. Nevertheless, here, we restrict ourselves to the $EB$ pair, which likely dominates the total signal-to-noise ratio (SNR) of the reconstruction, since $EB$ correlations should be zero in the absence of systematics. Its detection is thus a clearer indication of the $m-$bias field.

Finally, after computing $\hat{m}(\boldsymbol{L})$ in Fourier space, one can transform it back to real space, to obtain a two-dimensional map $\hat{m}(\boldsymbol{\theta})$ of the $m-$bias across the survey region. The resulting $\hat{m}$ map can be used to diagnose $m-$bias in cosmic shear survey data. In the next sections, we test this estimator on simulated Stage-IV-like shear maps, with injected artificial $m-$bias patterns, to validate its performance.

\section{Survey simulations and $m-$bias realisations}
\label{sec:realisations}

\subsection{The testing methodology}
\label{subsec:testing_method}
We aim to first test the method on simplified realisations of weak lensing surveys, quantifying the ability of the method to recover some injected patterns of $m-$bias variation, noting the point that the method relies on detection of mode-coupling, and hence is sensitive only to spatially-dependent bias.

In weak lensing surveys, the cosmic shear signal is buried within the "shape noise" caused by the broad distribution of intrinsic galaxy ellipticity (see e.g. \citealt{li_2023}). We show below that, for realistic galaxy number density, it is not possible to recover percent-level $m-$biases at arbitrary positions within a survey.  Any $m-$bias we measure has much larger statistical noise than the signatures that we aim to detect in Stage\,IV surveys.

We can overcome this limitation by analogous methods to those adopted for $c-$bias detection. First, if we have position-dependent templates or maps of systematic effects that are possible sources of bias, we may compare those maps with the $m-$bias map and assess the significance of any detection. Such systematics templates might be maps of PSF error, or perhaps other astronomical effects such as Galactic reddening, that might impact the shear measurement.

The second method is just to stack the $m-$bias estimator outputs themselves on some length-scale, analogously to the $c-$bias stacking methods. This approach is particularly appropriate for the case where $m-$bias arises from instrumental effects within the FOV - for example, some spatially-varying PSF error which is invariant in telescope coordinates, leading to a periodic bias variation across the survey, dependent on the set of instrument pointings that make up the survey.

We shall investigate both approaches in the following sections. In order to reproduce the effect of a replicating pattern, we generate multiple realisations of patches of survey data, with randomly-distributed galaxy shear values across the entire survey, but with an $m-$bias pattern that is invariant between the realisations.

\subsection{Galaxy survey sample}
\label{subsec:galaxy_sample}
  The cosmic shear signal power spectrum $C_\ell^{EE}$ is related to the matter power spectrum via
  \begin{equation}
    \label{eq:limber}
    C_\ell^{EE}=\int_0^{{\chi}_h}\frac{d\chi}{\chi^2} \left[W^{E}(\chi)\right]^2 P_{mm}\left(k=\frac{\ell+1/2}{\chi}, \chi\right) \, ,
  \end{equation}
  where $\chi(z)$ is the comoving distance for flat cosmologies, $\chi_h$ the comoving distance at the horizon and $P_{mm}(k,\chi)$ is the matter power spectrum as a function of wavenumber $k$ and comoving distance (which we use here as a proxy for redshift, or cosmic time, in the lightcone). The weak lensing kernel is
  \begin{equation}
    \label{eq:kernel}
    W^E(\chi)=\frac{3\Omega_{m,0}H_0^2}{2c^2}\chi[1+z(\chi)]\int_{z(\chi)}^{\infty}dz\,p(z)\frac{\chi(z)-\chi}{\chi(z)}, 
  \end{equation}
  with $\Omega_{m,0}$ the fraction of the matter density of the Universe at present, $H_0$ the Hubble constant, $c$ the speed of light and $p(z)$ the galaxy sample's redshift distribution. The expressions above make use of the Limber approximation \citep{Limber1953,1992ApJ...388..272K}, appropriate for the broad weak lensing kernel.

  To calculate $C_\ell^{EE}$ we use the Core Cosmology Library \citep[{\tt CCL},][]{Chisari_2019}, with the matter power spectrum given by the \texttt{HALOFIT} parameterisation of \citet{Smith2003} and \citet{Takahashi2012}. We consider a \Euclid-like \citep{mellier_2025} redshift distribution describing the photometric/cosmic shear sample with the form
  \begin{equation}\label{eq:n_of_z}
    p(z)\propto z^\alpha \exp\left[-\left(\frac{z}{z_0}\right)^\beta\right] \, ,
  \end{equation}
  with $\alpha=2$, $\beta=1.5$ and $z_0=0.637$ in $0<z\leq3$ \citep{laureijs2011Eucliddefinitionstudyreport}. We assume a total number density of $\bar{p}_\text{gal}=30\ \text{arcmin}^{-2}$ (see left panel of Fig.\ref{fig:n_z_and_m_bias_input} for $p(z)$ normalised to unit integral). Most cosmological analyses will make use of tomography, dividing the sample into different redshift bins to enhance the resulting cosmological constraints. Instead here we simply assume a single redshift bin containing the whole sample. On the one hand, this is the simplest approach to showcase the potential of the method as a proof of concept. On the other, combining the full sample reduces the level of shape noise in the data, which likely improves the sensitivity to a redshift-independent (but spatially-varying) $m$-bias. 
  
  The contribution to the total observed power spectrum receives a contribution from shape noise of the form
  \begin{equation}\label{eq:shape_noise}
    N_\ell^{EE}=N_\ell^{BB}=\sigma_e^2/\bar{p}_{\rm gal},
  \end{equation}
  with $\sigma_e=0.28$ the intrinsic ellipticity variance for a \Euclid-like galaxy sample \citep{mellier_2025}.
  
  Besides the contribution from gravitational lensing, the correlated galaxy shapes also receive a contribution from the intrinsic alignments (IA) of galaxies due to the tidal interactions and other long-range astrophysical effects taking place during galaxy formation \citep{Joachimi_2015}. In the simplest parameterisation, the so-called non-linear alignment model \citep[NLA,][]{Bridle_2007}, the IA contribution is a pure $E$ mode, and its presence therefore does not affect the performance of the quadratic estimator to reconstruct the $m$-bias map. Furthermore, omitting the IA contribution to $C_\ell^{EE}$ in the quadratic estimator does not significantly affect our results. We verified this by including this contribution, choosing an IA amplitude $A_{\rm IA}=2$, both in $C_\ell^{EE}$ and in the simulated maps from the Gaussian realisations (for more details, see Sec \ref{seubsec:gaussian}) used in the tests described below, or including it only in one of the two. In both cases, the reconstructed $m$-bias maps were indistinguishable from the maps reconstructed in the absence of IAs.
  
  In addition to IA, weak lensing measurements are also affected by baryonic feedback, which systematically suppresses the matter power spectrum at small scales \citep{Chisari_2019b}. For this, we performed the same series of tests using the \texttt{HMCode} \citep{Mead_2021}, with the baryonic feedback efficiency parameter $\log_{10}(\rm T_{\rm AGN}/\rm K)$ set to 7.8, which is the fit to the $\nu\Lambda$CDM BAHAMAS simulations, and also found that it did not affect the results. Thus, we decided to not include IA or baryonic effects in the main analysis in Sec.\,\ref{sec:results}.

\subsection{$m-$bias patterns}
\label{subsec:patterns}
We introduce two "toy models" describing flat-sky spatially-varying $m-$bias patterns $m(\boldsymbol{\theta})$, inspired by the models applied in \cite{kitching_2019}, which could be induced by residual effects in the PSF (see their Sec.\,3). We define a square patch of size $L_x=L_y$ (in radians) sampled on an $N_x\times N_y$ pixel grid. Let $\boldsymbol\theta=(\theta_x,\theta_y)$ denote angular coordinates in radians, with grid arrays $X,Y$ covering $[-L_x/2,L_x/2]\times[-L_y/2,L_y/2]$. We introduce a ``blob''-like pattern by parameterising $m(\boldsymbol{\theta})$ as:
\begin{equation}
    \label{eq:blob_mbias}
m_{\rm blob}(\boldsymbol\theta)= A \sin\big(k_{\rm b}|\theta_x|\big)\sin\big(k_{\rm b}|\theta_y|\big) \, ,
\end{equation}
with $A$ an arbitrary constant, and $k_{\rm b}$ is the angular frequency characterising the size of the blobs set to 50. An example can be seen in  the middle panel of Fig.\ref{fig:n_z_and_m_bias_input}. As an alternative example, we introduce a ``strip''-like pattern, given by
\begin{equation}\label{eq:strip_mbias}
  m_{\rm strip}(\boldsymbol\theta)= -A\sin\big(\pi-k_{\rm s}\theta_x\big),
\end{equation}
where $k_s$ characterises the size of the strips along the $x$ axis set to 100 (see right panel of Fig.\ref{fig:n_z_and_m_bias_input}).

In both cases we fix the amplitude parameter $A$ to ensure a desired rms variation $m_{\rm rms}$
\begin{equation}
  m_{\rm rms}=\sqrt{\left\langle m^2_{\rm pattern}\right\rangle_{\rm pix}},
\end{equation}
where $\langle\cdot\rangle_{\rm pix}$ denotes the pixel average over the $N_x \times N_y$ grid. We consider two rms values, $m_{\rm rms}=0.05$ (5\%) and 0.01 (1\%), motivated by the values
identified by \citet{cragg_2023} as being potentially of significance for Stage\,IV weak lensing surveys.

\subsection{Gaussian realisations}
\label{seubsec:gaussian}
The $m-$bias modulates the cosmic shear signal as well as the intrinsic shape noise, both of which we also simulate. As a first simple model, we generate Gaussian realisations of both components.

We generate flat-sky signal realisations following the $E$-mode power spectrum in Eq. (\ref{eq:limber}), using the \texttt{NaMaster} package \citep{Alonso_2019}, including angular scales in the range $[\ell_{\rm min},\,\ell_{\rm max}]$, with $\ell_{\rm min}=2\pi/L_{\rm side}$ and $\ell_{\rm max}=\sqrt{2}\pi N_{\rm side}/L_{\rm side}$. The input power spectrum was generated for a fiducial $\Lambda$CDM cosmology with parameters
\begin{equation}
  (\Omega_c,\Omega_b,h,\sigma_8,n_s)=(0.2432,0.0473,0.6898,0.8364,0.969),
\end{equation}
where $\Omega_c$ is the present cold dark matter density fraction, $\Omega_b$ is the present baryon density fraction, $h=H_0/100 \,\mathrm{km\,s^{-1}\,Mpc^{-1}}$ is the dimensionless Hubble constant, $\sigma_8$ is the rms of the linear matter density fluctuations in spheres of 8$h^{-1}$Mpc, and $n_s$ is the scalar spectral index.

We assign shape noise in each pixel as follows. For simplicity we assume the same number of galaxies in each pixel, given by $N_p=\bar{p}_{\rm gal}L_xL_y$. We then generate values for the shape noise in each pixel and shear component as uncorrelated Gaussian numbers with zero mean and a standard deviation $\sigma_p=\sigma_e/\sqrt{N_p}$.

Finally, the observed shear field is generated by adding the signal and noise components, and multiplying them by the $m-$bias patterns described in the previous section. These are then used in Eq.\,\ref{eq:estimator_def} to reconstruct the input bias pattern.

\subsection{N-body simulations}
\label{subsec:n-body}

Apart from using Gaussian realisations, we also test our quadratic estimator with N-body simulations from the \textit{cosmo}-SLICS suite \citep{Harnois_D_raps_2019}. The output of these N-body runs are convergence $\kappa$ and shear $\gamma_1, \gamma_2$ maps derived with 26 $w$CDM cosmology models and two random seeds each. In addition, for each model, there is a light-cone spanning from $z_\text{min}=0$ to $z_\text{max}=3$ with fixed redshift snapshots, spanning from 15 to 28 depending on the cosmology model. There are 50 light-cones per cosmology model, in total. We choose their fiducial $\Lambda$CDM cosmology model maps (see Sec.\,\ref{seubsec:gaussian} for the exact values). The shear maps $\gamma_1, \gamma_2$ are grids of $7745\times 7745$ pixels spanning an area of $100\deg^2$ ($10\deg$ per grid side). Based on these, we construct a ray-tracing of each light-cone given the \Euclid-like $p(z)$ redshift distribution (see Eq.\,\ref{eq:n_of_z}) by estimating the weighted sum of the shear maps $\sum_i w_i \gamma_{1}^i$, $\sum_i w_i \gamma_{2}^i$, where $w_i=p(z_i)\Delta z_i$, with $i$ the redshift snapshot and $\Delta z_i$ their half intervals. Then, we assign the shape noise and the $m-$bias pattern in the pixel grids that enter the quadratic estimator, as we similarly described for the Gaussian realisations in Sec.\,\ref{seubsec:gaussian}. Although the N-body simulations have a limit to their available volume, the Gaussian realisations contain only two-point information, so they cannot capture the non-linear, non-Gaussian structure of the shear field that arises from late-time structure formation. The \textit{cosmo}-SLICS N-body light-cones deliver the non-Gaussian lensing maps with realistic small-scale power and super-sample coupling. Therefore, these simulations are more accurate descriptions of the observed cosmic shear field. In addition, non-linearity can lead to coupling between $\ell$ modes, which would act as an additional source of noise for the reconstructed $m-$bias.

\begin{figure}
    \centering
    \includegraphics[width=0.9\linewidth]{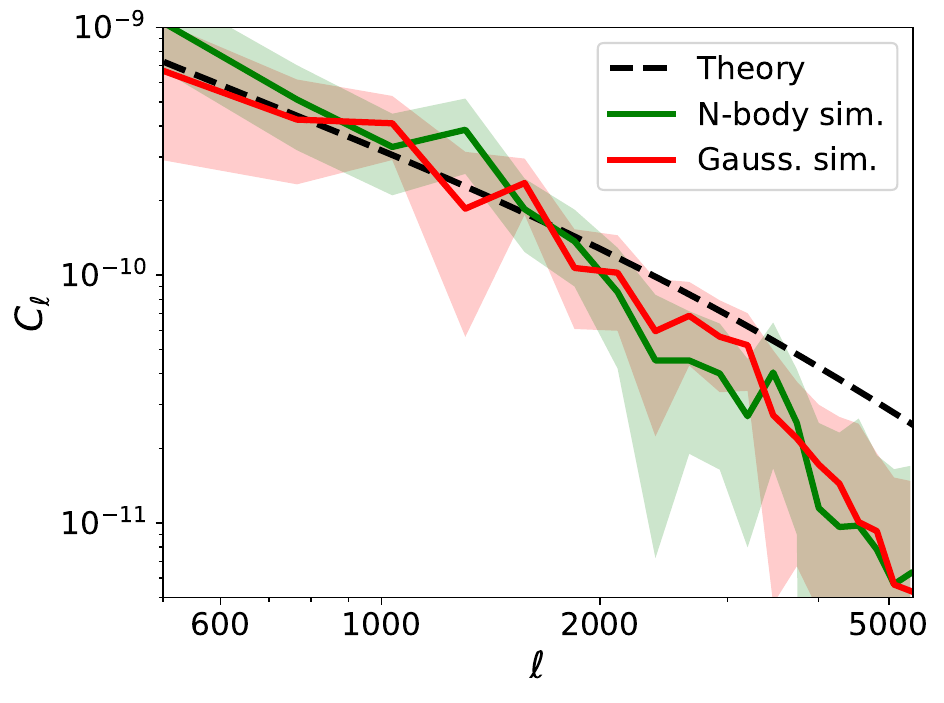}
    \caption{The angular power spectra from an N-body (green) and a Gaussian (red) simulation at an area of $1\deg^2$ with their $1\sigma$ uncertainties and the input theory prediction (dashed black). The high-$\ell$ suppression visible in the figure is due to finite-resolution effects.}
    \label{fig:Nbody_and_Gauss_cls}
\end{figure}

Despite the lack of non-linear structures in the Gaussian realisations compared to the N-body simulations at small scales below $1\deg^2$, the difference between their angular power spectra is expected to be small (see Fig.4 of \cite{Harnois_D_raps_2019}). In Fig.\,\ref{fig:Nbody_and_Gauss_cls}, we make a quick sanity check by comparing one N-body simulation with one Gaussian realisation. We compute the flat-sky angular power spectra along with their $1\sigma$ uncertainty bounds using the \texttt{pymaster} package for one random pair of $\gamma_1$ and $\gamma_2$ pixel maps ($31\times31$) from one N-body simulation and a Gaussian realisation  (green and red lines). We downsample (more details on the downsampling is described in Secs\,\ref{subsec:100degsq} and \ref{subsec:1degsq}) the Gaussian realisation here the same way we do for the N-body simulations, in order to have the same resolution and pixel window size effects. We also plot the theory prediction from \texttt{HALOFIT}. Indeed, we found that the two spectra are in agreement within their $1\sigma$ uncertainties. However, we see that the simulated spectra start to deviate from the theory prediction at $\ell\gtrsim2000$, due to resolution and pixel window effects (see \cite{Harnois_D_raps_2019}).
We have validated that, assuming a theory power spectrum in the estimator of Eq.\,\ref{eq:estimator_def}, described by a simple \texttt{HALOFIT} at up to $\ell_{\rm max}\lesssim 1000$, in order to avoid such deviations, we are able to reconstruct accurately the non-Gaussian $m-$bias patterns without any loss of information.

\section{Results}
\label{sec:results}

\subsection{Evaluation of results}
As discussed in Sec.\ref{subsec:testing_method}, an average over a number of realisations of the integral of Eq.\ref{eq:estimator_def},
equivalent to averaging over multiple sky patches in a weak lensing survey, is needed in order to suppress the shape noise. 

Although, in reality, the $m-$bias pattern and its rms amplitude is generally unknown, in cases where there are available templates describing certain types of systematic effects such as the PSF model, these could be compared to the data in a $\chi^2$ analysis. In this exercise, we have created $m-$bias toy models that we assume as templates (see again Eqs\,\ref{eq:blob_mbias}-\ref{eq:strip_mbias}). Therefore, we can quantify the confidence with which the quadratic estimator in Eq.\ref{eq:estimator_def} is able to reconstruct the input $m-$bias pattern in terms of its spatial variation and rms, by defining a SNR. This way, we can come up with an estimate of the number of realisations or sky-patch stacks one would need in order to detect an $m-$bias pattern of a certain rms. An estimate of the SNR even in the absence of a template is also possible by simply measuring the detection significance of the signal itself compared to noise.

\subsection{Signal-to-noise estimates}\label{subsec:snr}
To assess the performance of the quadratic estimator, we must quantify the significance of the reconstructed $m-$bias map. Let $\widehat{m}_i(\boldsymbol{\theta})$ be the reconstructed multiplicative bias map in the $i$-th simulated realisation, where each realisation corresponds to a sky patch of a large-area survey. We estimate the local variance of a given realisation as
\begin{eqnarray}
  s^2(\boldsymbol{\theta})=\frac{1}{N-1}\sum_{i=1}^N\left[\widehat{m}_i(\boldsymbol{\theta})-\bar{m}_N(\boldsymbol{\theta})\right]^2,
\end{eqnarray}
where $N$ is the total number of realisations, and $\bar{m}_N$ is the mean reconstructed map
\begin{eqnarray}
  \bar{m}_N(\boldsymbol{\theta})=\frac{1}{N}\sum_{i=1}^N\widehat{m}_i(\boldsymbol{\theta}).
\end{eqnarray}
Additionally, since all realisations are independent, the standard deviation of $\bar{m}_N$ is $\sigma_{\bar{m}}(\boldsymbol{\theta})\equiv s(\boldsymbol{\theta})/\sqrt{N}$.

\begin{figure}
    \centering
    \includegraphics[width=0.9\linewidth]{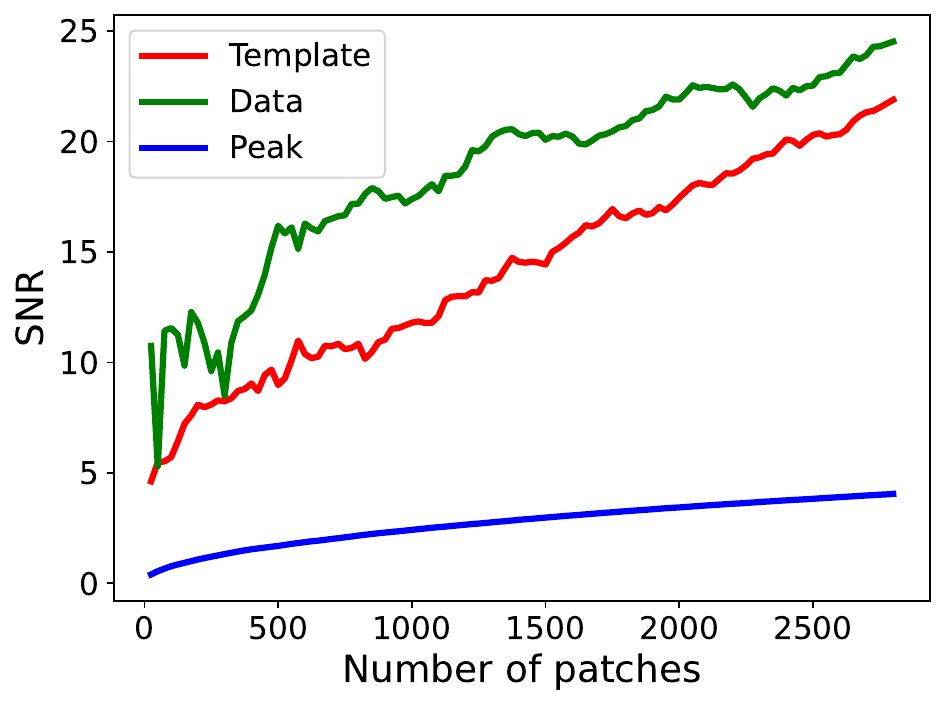}
    \caption{The 'template', 'data' and 'peak' SNR definitions as a function of the number of patches. We consider patches of $10\times10\deg^2$ and a 5\% rms blob-like $m-$bias.}
    \label{fig:SNRs}
\end{figure}

We use three different SNR  metrics to quantify the significance of the reconstructed $m-$bias map:
\begin{itemize}
  \item {\bf Template fitting SNR:} let us assume a linear model for the reconstructed $m-$bias map of the form ${\bf d}=\alpha {\bf t}+{\bf n}$, where ${\bf d}$ is a vector containing the reconstructed map in each pixel, and ${\bf t}$ is a known template map (e.g. the focal-plane map of a given systematic known to be a potential source of multiplicative bias), with $\alpha$ an unknown constant, and ${\bf n}$ is the reconstruction noise component. The significance of the detection in this case may be estimated as
  \begin{equation}
    {\rm SNR}_{\rm temp}=\bar{\alpha}/\sigma_\alpha,
  \end{equation}
  where $\bar{\alpha}$ and $\sigma_\alpha$ are the best-fit value of $\alpha$ and its statistical uncertainty, given by
  \begin{equation}
    \label{eq:a_and_sigma}
    \bar{\alpha}=\frac{\mathbf t^{\rm T} C^{-1} \mathbf d}{\mathbf t^{\rm T} C^{-1} \mathbf t}, \qquad \sigma_\alpha=\frac{1}{\sqrt{\mathbf t^{\rm T} {\sf C}^{-1} \mathbf t}}.
  \end{equation}
  Here ${\sf C}$ is the per-pixel covariance matrix, which we approximate as ${\sf C}={\rm diag}(\sigma_{\bar{m}}^2)$.
  \item {\bf Data-driven SNR:} one can also derive a SNR estimate from the data itself. The null $\chi^2$ is defined as
  \begin{equation}
    \label{eq:null_chi2}
    \chi^2_{\rm null}=\mathbf d^{\rm T} C^{-1} \mathbf d,
  \end{equation}
  and by considering the degrees of freedom, which in our case is the number of pixels $N_{\rm pix}$, we can derive a $\sigma$-detection definition which is
  \begin{equation}
    \label{eq:SNR_data}
    \rm{SNR}_{\rm data}=\sqrt{\chi^2_{\rm null}-N_{\rm pix}}.
  \end{equation}
In case of no $m-$bias in the data, $\chi_{\rm null}^2\rightarrow N_{\rm pix}$, so that $\rm{SNR}_{\rm data}\rightarrow 0$
  \item {\bf Peak SNR:} finally, for an alternative SNR definition, one can define a “peak mask’’ $\mathcal S$, selecting the brightest parts of the injected pattern, specifically those pixels where the absolute value of the true $m-$ bias map is above half of its maximum value
  \begin{equation}
    {\cal S} \equiv\left\{\boldsymbol\theta: |m(\boldsymbol\theta)|>\tfrac{1}{2}\max_{\boldsymbol\theta'} |m(\boldsymbol\theta')|\right\}.
  \end{equation}
  We then average the signal amplitude of the true map and the noise from the realisations over this mask and form a local SNR dubbed as ``peak'':
  \begin{equation}\label{eq:SNR}
    \mathrm{SNR}_\text{peak}=\frac{\left\langle |m(\boldsymbol\theta)|\right\rangle_{\boldsymbol\theta\in\mathcal S}}{\left\langle \sigma_{\widehat m}(\boldsymbol\theta)\right\rangle_{\boldsymbol\theta\in\mathcal S}}.
  \end{equation}
  This SNR definition gives a sense of the significance with which the largest deviations in the $m-$bias pattern may be identified in a single pixel.
\end{itemize}

\begin{figure*}
    \centering
    \includegraphics[width=1.0\linewidth]{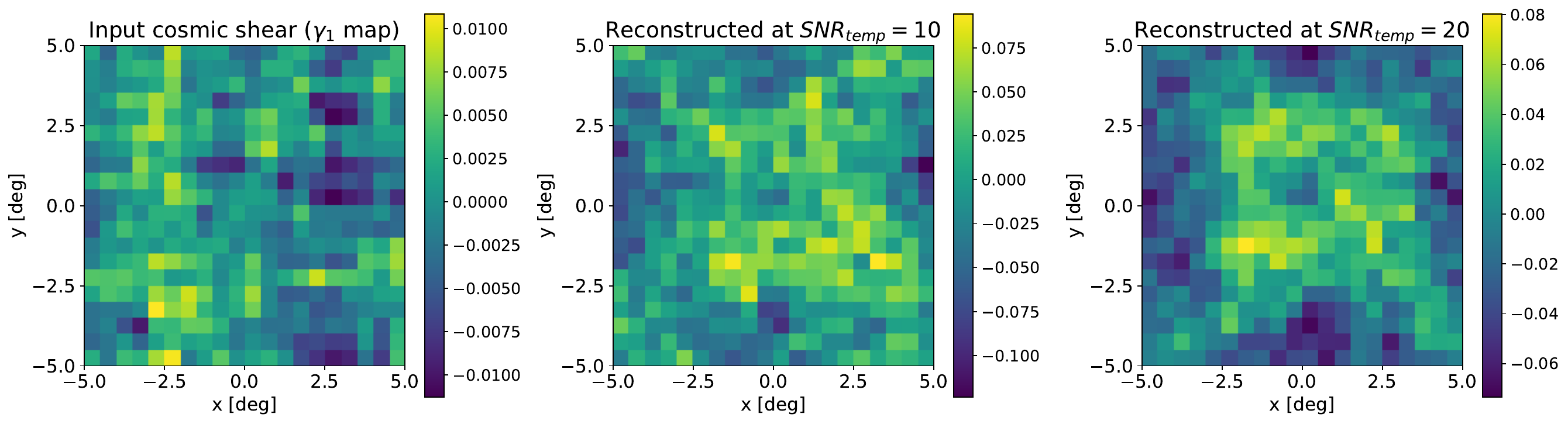}
    \includegraphics[width=1.0\linewidth]{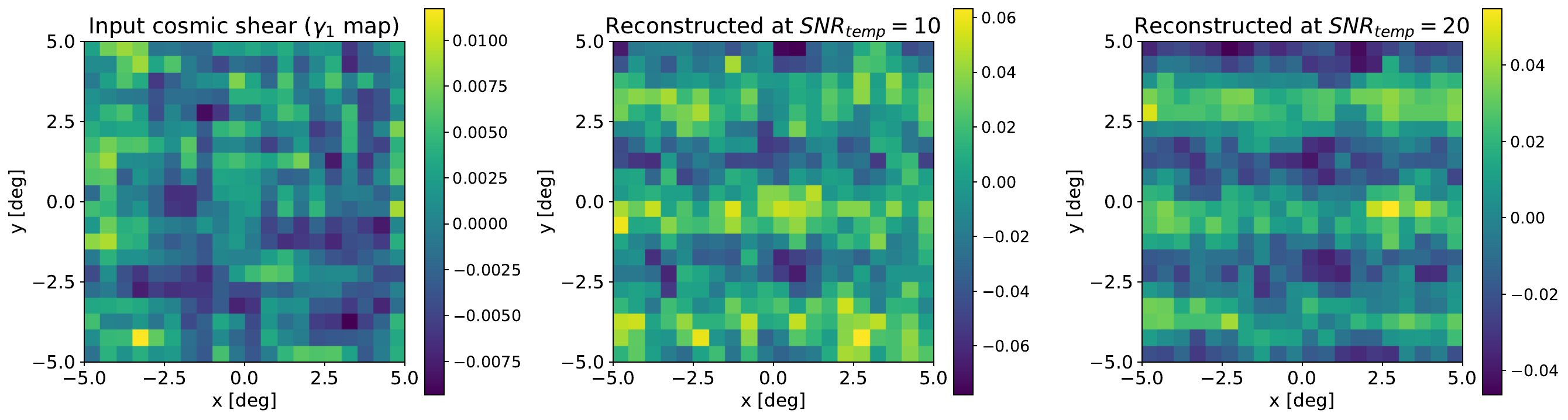}
    \caption{\textit{Top left:} One input $\gamma_1$ component of the cosmic shear from Gaussian realisations. \textit{Top middle}: Reconstruction of the 5\% rms blob-like $m-$bias field at $\rm SNR_{\rm temp}$=10 with 600 realisations. \textit{Top right}: Reconstruction of the same $m-$bias field at $\rm SNR_{\rm temp}$=20 with 2800 realisations. The grid area is 100 $\deg^2$ with $20\times20$ pixels. Similarly, the bottom panels show the 5\% rms strip-like $m-$bias model reconstructed with 2000 (bottom middle) and 6700 realisations (bottom right).}
    \label{fig:mbias_10_and_20_sigma_Gauss_100deg2}
\end{figure*}

We established the number of sky patches required to reach a target $\mathrm{SNR}_{\rm def}^{\rm req}$, by increasing the number of patches until $\mathrm{SNR}_{\rm def}\geq \mathrm{SNR}_{\rm def}^{\rm req}$. 

\subsection{$m-$bias reconstruction on large scales}
\label{subsec:100degsq}


We aim to test the $m-$bias quadratic estimator at large-scales. We start by constructing Gaussian realisations of $10\times10\deg^2$ patches. We choose a $N_x\times N_y=20\times20$ pixel grid for our maps and the number of galaxies in each patch is $\sim10^7$. In principle, one could choose a finer pixel grid, but the reason not to opt for this is two-fold. 
First, the spatially-varying $m-$bias toy models we explore here are characterized by large enough scale variations corresponding to 5 deg (blob-like) and 2 deg (strip-like) in a 100 deg$^2$ area, which are captured with such a pixel grid size (analogously, the same is true for the $1\deg^2$ patches we investigate in Sec.\ref{subsec:1degsq}). Secondly, it is a considerable speed-up for the code as well, since the computational cost scales with the FFT done on the number of pixels in the grid, allowing us to explore large numbers of realisations.



\begin{figure*}
    \centering
    \includegraphics[width=0.8\linewidth]{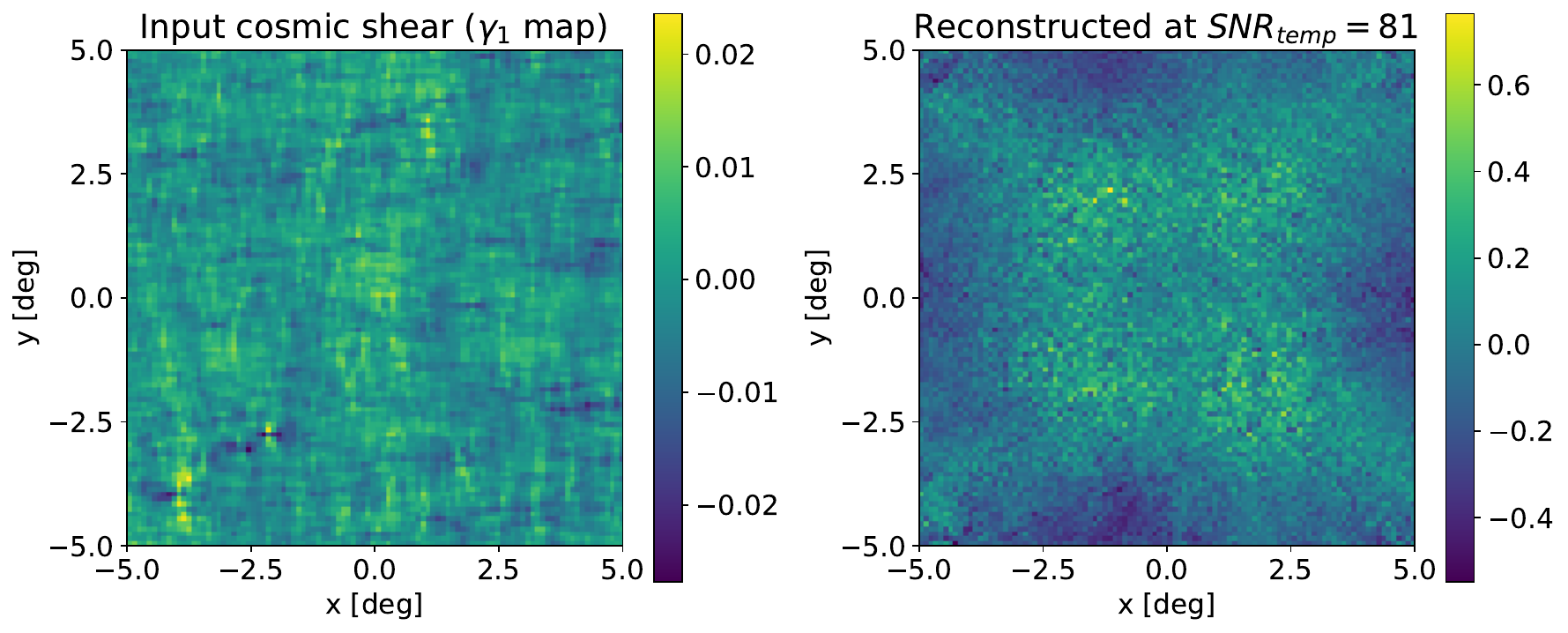}
    \includegraphics[width=0.8\linewidth]{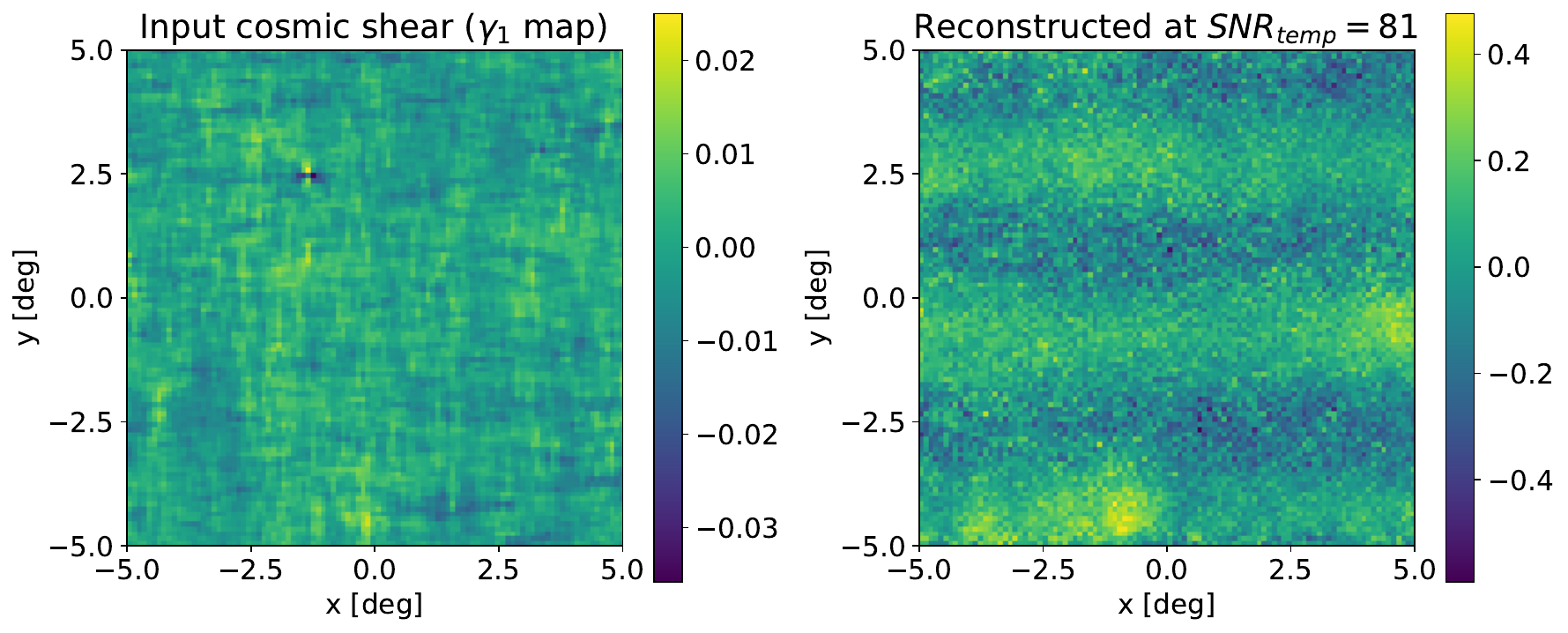}
    \caption{\textit{Top left:} One input $\gamma_1$ component of the cosmic shear from N-body realisations. \textit{Top right}: Reconstruction of the 20\% rms $m-$bias field at $\rm SNR_{\rm temp}$=81 with 50 simulations. The grid area is 100\,$\deg^2$, with $100\times100$ pixels. Bottom panels are for a 20\% rms strip-like $m-$bias, reconstructed with $\rm SNR_{\rm temp}$=81, using also 50 simulations.}
    \label{fig:mbias_Nbody_100deg2}
\end{figure*}

Within this setup, and using a blob-like $m-$bias template with 5\% rms fluctuations, Fig.\ref{fig:SNRs} shows the number of stacked realisations needed to achieve a given SNR values with the ``template'', ``data'' and ``peak'' definitions. The ``peak'' definition is naturally stricter, since it quantifies the SNR in a single bright pixel. It reaches a value of ${\rm SNR}_{\rm peak}=4$ for 2800 realisations, at which point the template and data-based definitions reach ${\rm SNR}_{\rm temp}\sim20$ and ${\rm SNR}_{\rm data}\sim24$. Fig. \ref{fig:mbias_10_and_20_sigma_Gauss_100deg2} shows one of the Gaussian realisations used for this exercise, as well as the reconstructed $m-$bias map after stacking enough realisations to achieve a total ${\rm SNR}_{\rm temp}=10$ (corresponding to ${\rm SNR}_{\rm peak}=2$), where the bias pattern starts becoming visually evident. We can also appreciate that the reconstruction looks less noisy when we reach ${\rm SNR}_{\rm temp}=20$.
For a statistical detection of the $m-$bias through template fitting, corresponding to ${\rm SNR}_{\rm temp}=5$ (i.e. a ``$5\sigma$'' detection), we require 50 realisations, corresponding to a total sky area of 5000 square degrees (even though the $m-$bias pattern is not visually evident at that point). The quadratic estimator approach proposed here may therefore be used to identify a spatially-varying multiplicative bias as long as an informed template for its spatial pattern can be constructed. This may be constructed, for example, as sky maps of scalar survey conditions known to be highly correlated with shape measurement errors (e.g. co-added PSF size, or Galactic reddening, which may impact shear measurement through its systematic effect on galaxy observed spectral energy distributions). This procedure is reminiscent of the techniques used to identify additive systematics in cosmic shear and galaxy clustering \citep{Berlfein_2024}, but now applicable to multiplicative bias effects. 

The bottom panels in Fig. \ref{fig:mbias_10_and_20_sigma_Gauss_100deg2} show the same results for a strip-like $m-$bias, where the same conclusions drawn for the blob-like case hold qualitatively (6700 realisations are now needed). In what follows, we report the number of realisations/patches needed to achieve a ${\rm SNR}_{\rm temp}=20$ unless otherwise stated, where the $m-$bias pattern can be visually identified (corresponding to ${\rm SNR}_{\rm peak}\sim5$ and ${\rm SNR}_{\rm data}\sim24$ or much higher, depending on the case).

For independent realisations the reconstruction noise is the same and scales with the number of realisations $N$ as $\propto1/\sqrt{N}$.Therefore, a desired SNR is reached at $\propto {\rm rms}\sqrt{N}$ and going from $5\%$ to $1\%$ rms values requires $\sim25$ times the number of realisations to achieve the same SNR. Indeed, we recover the input 1\% rms $m-$bias at $\text{SNR}_{\rm temp}=20$, but now with an increased number of realisations by that factor, corresponding to 80750 and 160875 for the blob and the strip patterns, respectively. We acknowledge that these numbers of realisations or equivalently multiple sky patches needed imply large sky areas which are not available in a real data analysis. We only choose these numbers here as a reference in order to obtain the good visual reconstruction at $\text{SNR}_{\rm temp}=20$. In fact, we could still detect the $m-$bias of both patterns at $\text{SNR}_{\rm temp}=3$ with roughly 100 realisations, numbers certainly available in the upcoming data since they correspond to a total sky area of 10000 square degrees. The results for $\text{SNR}_{\rm temp}=20$ visually look very similar to the recovered patterns of Fig.\ref{fig:mbias_10_and_20_sigma_Gauss_100deg2}, and therefore we do not show them. It is worth noting here that the increased number of realisations needed is expected, given that we need to suppress the shear noise further in order to detect an $m-$bias with lower rms values.

We now turn our attention to the shear maps from the N-body simulations. As discussed in Sec \ref{subsec:n-body}, the N-body simulations include realistic non-Gaussianities that can affect the performance of the quadratic estimator, by introducing additional noise or biases in the $m$-bias reconstruction. The output maps are pixel grids of size $7745\times 7745$. We assign the shape noise and then downsample these maps with a Gaussian filtering, with a width corresponding to the new pixel Nyquist frequency and also with a cubic interpolation scheme, into grids of size $100\times100$. Afterwards, we multiply with the $m-$bias toy models like we did for the Gaussian realisations. The total number of realisations we consider here is 50 (25 light-cones generated with 2 random seeds). Given that small number of simulations, we examine an exaggerated $m-$bias scenario in order to have a good visual emergence of the $m-$bias field. Thus, by setting the rms to $20\%$ just for this case, we evaluate the performance of the quadratic estimator using all the 50 N-body simulations shear maps covering 100 $\text{deg}^2$ patches. In Fig. \ref{fig:mbias_Nbody_100deg2}, we see that with all the simulations available we manage to capture very well visually the spatial-variation of the input $m-$bias patterns at $\rm SNR_{\rm temp}\sim81$ for both the blob and strip models as seen in the top and the bottom panels of Fig. \ref{fig:mbias_Nbody_100deg2}, respectively ($\rm SNR_{\rm temp}\sim20$ is reached with only 5). At this point, we cannot safely conclude yet that our quadratic estimator is not affected by the non-linearities in the N-body simulations since we use an exaggerated 20\% rms $m-$bias, but this is a first test verifying that the quadratic estimator works using the N-body simulations. We will see in detail in Sec. \ref{subsec:1degsq} that indeed our estimator is also not impacted by these non-Gaussianities at small scales using realistic rms $m-$bias values. 


\begin{figure*}
    \centering
    \includegraphics[width=0.8\linewidth]{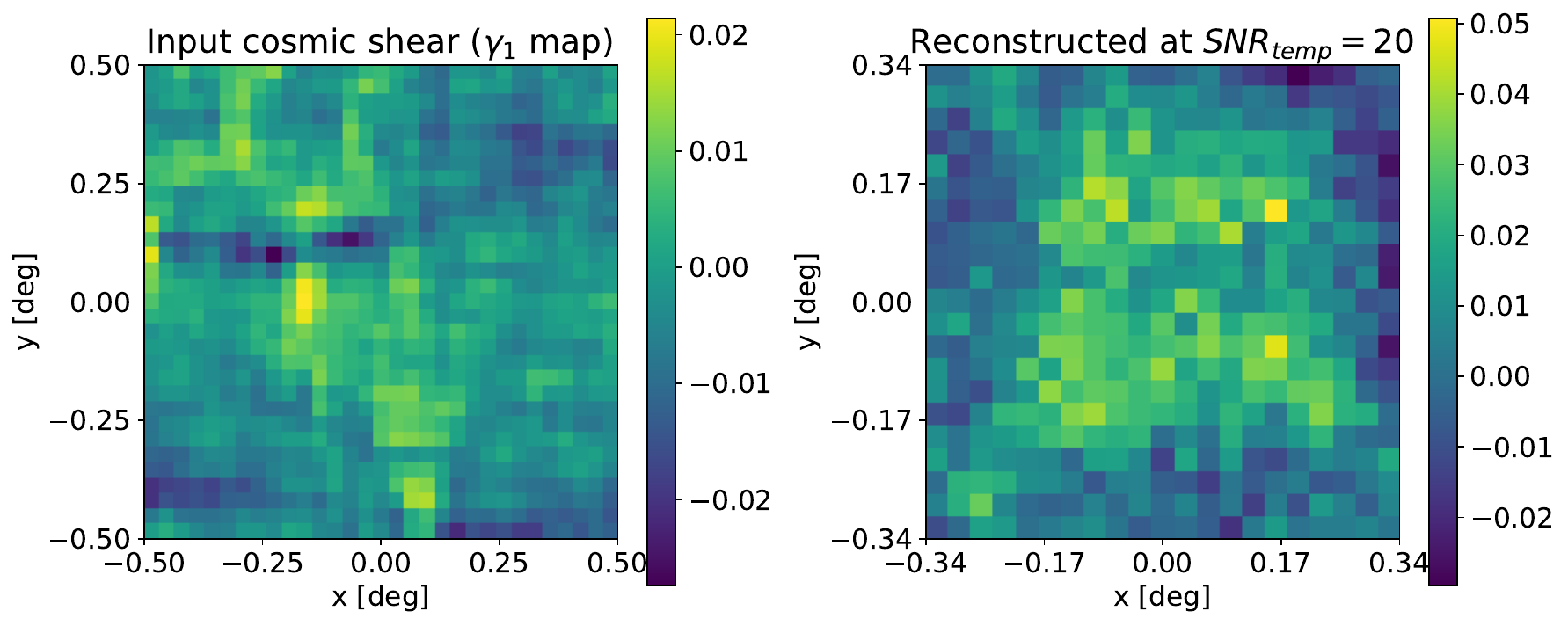}
    \includegraphics[width=0.8\linewidth]{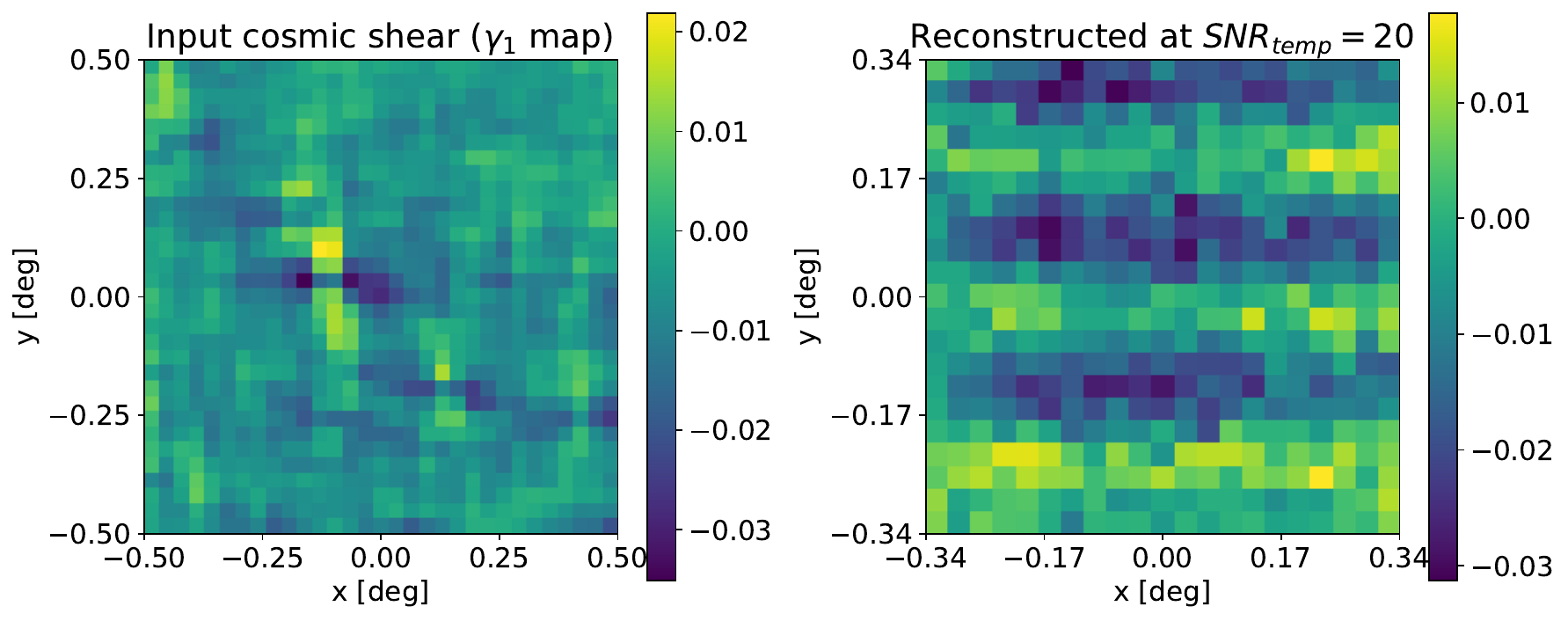}
    \caption{\textit{Top left:} One $\gamma_1$ component patch of the cosmic shear from N-body realisations. \textit{Top right}: Reconstruction of the $m-$bias field with $5\%$ amplitude at $\rm SNR_{\rm temp}=20$ with 400 patches. Bottom panels same as top but for a strip-like $m-$bias model with 1000 patches. The reconstructed grid area is  $0.68\times0.68\deg^2$  with $21\times21$ pixels.}
    \label{fig:mbias_Nbody_1deg2}
\end{figure*}

\subsection{$m-$bias reconstruction on small scales}
\label{subsec:1degsq}


There are good reasons to test the quadratic estimator at a smaller area of the order of $\sim1\times1\ {\rm deg}^2$: forthcoming weak lensing galaxy surveys, such as \Euclid \citep{mellier_2025} and LSST \citep{2012arXiv1211.0310L}, have $\sim$degree-sized fields of view (diameter $D_{\rm FOV}\sim0.7^\circ$ and $\sim3.5^\circ$, respectively). Many shape measurement systematic effects such as PSF modelling errors will be approximately constant in telescope coordinates, and therefore will leave a repeated (or at least highly-correlated) imprint across different exposures on scales smaller than $D_{\rm FOV}$. The presence of these systematics could thus be detected by stacking maps centred at those exposures in telescope coordinates using the quadratic estimator described here. The total sky area covered by these surveys, of order $\sim1.5\times10^4\,{\rm deg}^2$, would allow for $\sim1.5\times10^4$ degree-sized patches. Here we will show that these would be more than enough to detect a 5\% rms spatially-varying $m-$bias patterns at a high significance (e.g. $\rm SNR_{\rm temp}=20$) with a realistic small-scale cosmic shear signal from N-body simulations. Moreover, with a \Euclid-like redshift distribution, it would be possible to detect 1\% rms $m-$bias at the same significance with a few thousand patches.

For this test we consider the cosmic shear realisations from the N-body simulations, and assume the same \Euclid-like galaxy density number used above (see Sec.\,\ref{subsec:galaxy_sample}). The corresponding number of galaxies residing in a typical $1\,{\rm deg}^2$ patch is roughly $\sim10^5$ sources. In each of the 50 simulations, we split the $7745\times7745$ pixel grid spanning an area of $100\deg^2$ into 100 sub-grids of size $774\times774$, cropping 5 pixels at the outer edges to form a $10\times10$ tiling, each spanning an area now of $\sim1\deg^2$. Similarly to the procedure in Sec.\,\ref{subsec:100degsq}, we include Gaussian shape noise and downsample these grid maps with a Gaussian filter and a cubic interpolation scheme into $31\times31$ pixel grids. We imprint the multiplicative bias pattern in the central $21\times21$ pixels of each patch, leaving a buffer with a width of 5 pixels where $m(\boldsymbol{\theta})=0$. We do this to avoid edge effects caused by the implicit assumption of periodic boundary conditions in the FFTs. This assumption was valid in the previous sections, where we analysed the full simulation boxes, but it breaks down when we work with smaller shear patches: the shear field is discontinuous at the patch boundaries, so large-scale modes are poorly represented and generate artefacts. We therefore introduce a buffer region around each patch before convolving with the usual $m-$bias pattern and applying the quadratic estimator in Eq.\ref{eq:estimator_def}. We have verified that, without this buffer, spurious features appear in the reconstructed $m-$bias field near the patch edges, whereas they disappear when the buffer is included.

In Fig. \ref{fig:mbias_Nbody_1deg2}, we present the 5\% rms for blob-like and strip-like $m-$bias patterns reconstructed at $\rm SNR_{\rm temp}=20$ by averaging over 400 and 1000 patches. Since the appropriate templates at sub-FOV scales may be difficult to construct, we also report the data-driven $\rm SNR_{\rm data}$ values which are 30 and 39 using the 400 and 1000 patches, respectively. For an 1\% rms $m-$bias the $\rm SNR_{\rm temp}=20$ is achieved now with 2800 and 7600 patches for a blob-like and strip-like $m-$bias, respectively, corresponding to $\rm SNR_{\rm data}$ of 54 and 86. These constitute the largest detection significance reported here, and show that spatially-varying $m$-biases on sub-FOV scales with $\sim1\%$ amplitudes can be detected at high significance using the quadratic estimator approach. 

\subsection{Cosmology independence}
\label{subsec:cosm-indep}
  The quadratic estimator presented here requires an estimate of the cosmic shear power spectrum $C_\ell^{EE}$ (see Eq. \ref{eq:weight_function}). Since this is {\em a priori} unknown, it is interesting to quantify whether using an incorrect $C_\ell^{EE}$ would lead to a significant error in the reconstructed $m$-bias map. To do this, we repeat our analysis of the N-body simulation maps, which use the fiducial cosmology $\Lambda$CDM, but assuming a significantly different $w$CDM cosmological model to calculate the input $C_\ell^{EE}$, with parameters
  \begin{multline}
    (w, \Omega_c,\Omega_b,h,\sigma_8,n_s)=\\
    (-1.9127, 0.5009,0.0473,0.7645,0.4716,0.969).
  \end{multline}
  This cosmological model was shown in \cite{Harnois_D_raps_2019}, to exhibit strong differences in the shape of $C_\ell^{EE}$, due to the strong changes in the growth history and distance-redshift relation caused by a high $\Omega_c$ and an extreme phantom-like value of $w$. We use $1\deg^2$ patches, as in Sec.\,\ref{subsec:1degsq}. 
  
  Despite the inconsistent cosmology models between the simulations in the shear maps and the theory spectra, we manage to recover the $m-$bias field successfully. The reconstruction for a 1\% rms blob-like and strip-like $m-$bias with N-body simulations with ${\rm SNR_{\rm temp}}=20$ is achieved with $\sim$2800 and $\sim$7800 patches, respectively. The reconstructions for the 5\% $m-$bias patterns are similar to the findings in Sec. \ref{subsec:1degsq} as well. This means that the quadratic estimator formalism in Eq. \ref{eq:estimator_def} we employ here can be largely insensitive to the cosmology model we assume in the theory angular power spectrum entering the quadratic estimator. This is not entirely surprising: the estimator recovers the $m$-bias map by correlating $E$-$B$ Fourier modes with different wavenumbers. Using an incorrect $C_\ell^{EE}$ only leads to a sub-optimal weighting of the different mode pairs, which may affect the noise in the reconstructed $m$-bias map, but not the ability of the method to recover it.

\subsection{Additive bias contamination}\label{subsec:c-bias-contamination}
  We also test whether the presence of an additive $c$-bias affects our ability to reconstruct the $m$-bias. The $c$-bias is expected to be one order of magnitude smaller than the $m-$bias in Stage-IV galaxy surveys \citep{kitching_2019}. Thus, we consider the case of an rms strip-like $c-$bias pattern amplitude of 0.001 added to the shear field and then multiplied by a $1\%$ rms blob-like $m-$bias pattern (similar results are found when swapping the strip-like and blob-like patterns). We find that the presence of a $c$-bias contaminant has no discernible impact in the reconstructed $m$-bias map, and does not change any of our conclusions regarding its detectability.
  
  We further confirm that our quadratic estimator definition in Eq. \ref{eq:estimator_def} is only sensitive to the $m-$bias, and not to the $c-$bias, by setting the $m-$bias pattern to zero and keeping only the $c-$bias patterns. For this, we test $c-$bias maps with an rms amplitude of 0.001, as well as two unrealistic cases with much larger amplitudes of 0.01 and 0.05. In all cases, we fail to reconstruct the input $c-$bias patterns, and the reconstructed $m$-bias is compatible with zero. This finding confirms that the quadratic estimator is insensitive to contaminating $c-$bias.
 
\section{Conclusions}
  We have presented and validated a quadratic estimator that directly reconstructs spatially-varying $m-$ shear bias, by exploiting the $EB$ off-diagonal mode-couplings that such a field induces in cosmic shear maps. Building on this, our estimator combines quadratic combinations of $E$ and $B$ with optimal normalised weights to reconstruct a map of $m-$bias variations to first order, in close analogy with standard CMB lensing reconstruction methods \citep{Lewis_2006}. This statistic is by construction insensitive to a constant $m-$bias, and it does not respond to $c-$bias.

  To quantify the method's ability to detect a given $m$-bias pattern, we used three SNR definitions: ``template'', ``data'', and ``peak'' (where the pattern appears visually), with the ``template'' definition being the most relevant when an external template for the spatial pattern is available. We have explored two different scenarios in which such an estimator could be deployed.
  
  First, multiple systematic effects are naturally described in telescope coordinates (e.g. PSF variations across the focal plane), and will imprint a repeating pattern across different exposures on scales smaller than the instrument's FOV. A spatially-varying $m$-bias may then be reconstructed by applying our quadratic estimator to FOV-sized patches and then stacking the result over multiple telescope pointings. In this case, we have shown that assuming a \Euclid-like survey and using non-Gaussian N-body shear field maps from the \textit{cosmo}-SLICS suite, we would be able to detect a blob-like (strip-like) $m$-bias pattern with a $5\%$ amplitude with a signal-to-noise ratio of ${\rm SNR}_{\rm temp}=20$ after stacking over only 400 (1000) patches. In turn, 1\% variations could be detected at the same significance with 2800 (7600) patches, still within the ``few-thousand-patch'' regime, feasible for a wide-area Stage-IV experiment. Similar conclusions hold if using the ``data'' SNR definition instead.

  Secondly, we explored the possibility of reconstructing variations on larger scales, as could be caused by varying observing conditions throughout the experiment's observing campaign. In such a scenario, a template search for spatially-varying $m$-bias may still be feasible, by building template maps of the most likely sources of shape measurement systematics from per-exposure metadata. On $10\times10\deg^2$ patches, Gaussian shear realisations show that a 5\% rms $m-$bias pattern is recovered at $\rm SNR_{\rm temp}=20$ with 2800 and 6700 co-added sky patches, for ``blob-like'' and ``strip-like'' patterns, respectively. Although visual morphology requires high SNR, template-amplitude fitting achieves $5\sigma$ detections with as few as $\sim50$ patches. For 1\% rms, reaching the same $\rm SNR_{\rm temp}=20$ significance demands of the order of 100000 patches, for both patterns, consistent with the need to suppress noise much further at the percent level but unfeasible given the galaxy survey areas. Despite that, a $3\sigma$ detection is still feasible with $\sim$100 patches. Using again the N-body simulations in the same 100 $\deg^2$ geometry, we were able to test the estimator for a large input bias of 20\% and found that this could be detected at high $\rm SNR_{\rm temp}\simeq81$. We were not able to test the method on larger areas, owing to the limited simulation volume available, but a naive scaling by the square root of the observed area predicts that $\rm SNR_{\rm temp}\simeq5$ could be achieved for 1\% bias with a sky area of order 8000 $\deg^2$,
  broadly consistent with the results from the Gaussian simulations.



  A set of robustness tests supports the method’s applicability to real analyses. First, changing the theory spectrum used in the estimator weights from the fiducial $\Lambda$CDM to a deliberately extreme wCDM model leaves the reconstruction performance essentially unchanged, indicating that the estimator’s response is largely insensitive to moderate cosmology mis-specification in the filtering. Secondly, injecting $c-$bias at realistic levels does not bias the $m-$bias reconstruction nor increase the number of stacks required, and, in the absence of $m$, the quadratic estimator does not yield any false detection arising from contaminating additive bias. Thirdly, including intrinsic alignments or baryonic feedback either in the Gaussian realisation maps, the theory spectra, or both produced indistinguishable results from the IA-free and baryon feedback-free case at the scales and noise levels investigated. Finally, the flat-sky implementation used here is adequate for the patch sizes analysed, while a curved-sky generalisation with spin harmonics and $\text{Wigner}-3j$ couplings should be straightforward in a future development.

  Taken together, these results demonstrate that spatially-varying $m-$bias at the few-percent to percent level is detectable at high significance with realistic survey characteristics, provided one stacks a sufficient number of FOV-scale patches or, when available, leverages physically motivated templates (e.g. PSF size/error proxies, detector coordinates). For the \Euclid/LSST surveys, the available number of independent FOVs comfortably supports the stack sizes implied by our $1\deg^2$ forecasts. There is also another natural extension apart from the curved-sky formalism. The method can include masks, and also allow for a forward-modelling of the reconstructed $m-$bias maps at the power spectrum level, via a weight map coupling, in order to correct the data from the spatially-varying $m-$bias pattern.  

  In summary, the estimator introduced here provides a practical and robust route to detecting spatially-varying $m$ shear bias at the angular scales and amplitudes most relevant for Stage\,IV weak lensing cosmology. Its demonstrated performance on both Gaussian and fully non-Gaussian shear fields from N-body simulations, its weak dependence on cosmology assumptions, and its limited susceptibility to $c$-bias contamination, make it immediately useful for \Euclid- and LSST-era systematics control, both as a survey-scale diagnostic and as a quantitative ingredient in end-to-end mitigation pipelines.

\section*{Acknowledgements}
  KT is supported by the STFC grant ST/W000903/1 and by the European Structural and Investment Fund. KT further acknowledges support by the European Union’s Horizon Europe research and innovation program under the Marie Sklodowska-Curie COFUND Postdoctoral Programme grant agreement No.101081355- SMASH and from the Republic of Slovenia and the European Union from the European Regional Development Fund. Views and opinions expressed are however those of the authors only and do not necessarily reflect those of the European Union or European Research Executive Agency. Neither the European Union nor the granting authority can be held responsible for them. DA acknowledges support from the Beecroft Trust. We made extensive use of computational resources at the University of Oxford Department of Physics, funded by the John Fell Oxford University Press Research Fund.



\bibliographystyle{mnras}
\bibliography{main} 







\bsp	
\label{lastpage}
\end{document}